\pdfoutput=1

\documentclass[aps,prl,reprint,showpacs,superscriptaddress]{revtex4-1}

\usepackage{graphicx}
\usepackage{graphics}
\usepackage{amsmath}
\usepackage{amssymb}
\usepackage{amsfonts}
\usepackage{dcolumn}
\usepackage{dsfont}
\usepackage{latexsym}
\usepackage{rotating}
\usepackage{color}
\usepackage{latexsym}
\usepackage{bbm}
\usepackage{subfigure}
\usepackage{float}
\usepackage{epsfig}
\usepackage{epsf}
\usepackage{psfrag}
\usepackage{bm}
\usepackage{amsthm}
\usepackage{eucal}
\usepackage{mathrsfs}
\usepackage{url}
\usepackage{braket}
\usepackage{epstopdf}
\usepackage{color} 

\usepackage{hyperref}
\hypersetup{
colorlinks=true,final=true,
        linkcolor=blue,
        citecolor=blue,
        filecolor=blue,
        urlcolor=blue,
}
  
\begin{document}
  
\title{A complete description of the magnetic ground state in spinel vanadates}
  
\author{Jyoti Krishna}
\affiliation{ Department of Physics, Indian Institute of Technology Roorkee, Roorkee - 247667, Uttarakhand, India}
\author{N. Singh}
\affiliation{Max Born Institute for Nonlinear Optics and Short Pulse Spectroscopy, Max-Born-Strasse 2A, 12489 Berlin, Germany}
\author{S. Shallcross}
\affiliation{Lehrstuhl f\"ur Theoretische Festk\"orperphysik, Staudstr. 7-B2, 91058 Erlangen, Germany}
\author{J. K. Dewhurst}
\affiliation{Max-Planck-Institut f{\"u}r Mikrostrukturphysik, Weinberg 2, D-06120 Halle, Germany}
\email{dewhurst@mpi-halle.mpg.de}
\author{E. K. U. Gross}
\affiliation{Max-Planck-Institut f{\"u}r Mikrostrukturphysik, Weinberg 2, D-06120 Halle, Germany}
\author{T. Maitra}
\affiliation{ Department of Physics, Indian Institute of Technology Roorkee, Roorkee - 247667, Uttarakhand, India}
\author{S. Sharma}
\affiliation{Max Born Institute for Nonlinear Optics and Short Pulse Spectroscopy, Max-Born-Strasse 2A, 12489 Berlin, Germany}

\date{\today}
  
\begin{abstract}

Capturing the non-collinear magnetic ground state of the spinel vanadates AV$_2$O$_4$ (A= Mn, Fe and Co) remains an outstanding challenge for state-of-the-art \emph{ab-initio} methods. We demonstrate that both the non-collinear spin texture, as well as the magnitude of local moments, are captured by a single value of the on-site Hubbard $U$ of 2.7~eV in conjunction with the local spin density approximation (LSDA+$U$), provided the source term (i.e., magnetic monopole term) is removed from the exchange-correlation magnetic field ${\bf B}_{XC}$. We further demonstrate that the magnetic monopole structure in ${\bf B}_{XC}$ is highly sensitive to the value of $U$, to the extent that the interplay between on-site localization and local moment magnitude is qualitatively different depending on whether the source term is removed or not. This suggests that in treating strongly correlated magnetic materials within the LSDA+$U$ formalism, subtraction of the unphysical magnetic monopole term from the exchange-correlation magnetic field is essential to correctly treat the magnetic ground state.

\end{abstract}

\maketitle


The strongly correlated electron systems (SCES) derive their richness from competing and coexisting multiple long range orders (LROs) such as charge, magnetic, orbital order \cite{dagotto}.
 A strong interplay among various degrees of freedom 
(e.g. charge, spin, orbital and lattice) in these materials provides a perfect platform for both basic and applied physics questions \cite{tokura1}. 
The family of spinel vanadates (AV$_{2}$O$_{4}$) belongs to such a class of materials where strong correlation, complex spin texture, and geometric frustration of the underlying lattice
work in tandem \cite{radaelli,slee}, and this richness of physics has attracted the sustained attention of the condensed matter and materials science community \cite{garlea,sarkar,slee,qzhang,elisa,suzuki}.

However, as several invesigations have made clear, the modern day theoretical method of choice, namely density functional theory (DFT) \cite{hohen,kohn}, fails to describe  spinel vanadates in two crucial ways. Firstly, in experiment the moment on V atoms is much lower (e.g. 1.3 $\mu_B$ in MnV$_2$O$_4$ \cite{garlea} to 0.65 $\mu_B$ in ZnV$_2$O$_4$ \cite{reehuis}) than the DFT values \cite{nanguneri,sarkar,ssarkar,tm} with difference between the two as high as 60\%. The reasons behind this large reduction in V moment remains contested with speculations including spin frustration, quantum fluctuations and spin-orbit interaction effects. Secondly, the experimentally observed ground state magnetic structure is a complex spin texture i.e. a non-collinear arrangement of V spins (with a large angle between A and V spins), while DFT predicts a collinear ferrimagnetic ground-state \cite{sarkar,raman,pardo,dibyendu,ssarkar,nanguneri,tm}. This incorrect DFT ground-state entails that the interesting physics of magnetic phase transitions in these materials stays beyond any \textit{ab-initio} description.

In the present work, with an example of three spinel vanadates (FeV$_2$O$_4$, MnV$_2$O$_4$ and CoV$_2$O$_4$), we probe the reason behind
the failure of DFT, which is otherwise an excellent theory for \emph{ab-initio} description of complex magnets. DFT is in principle an exact theory, but in practice requires an approximation for the so called  exchange-correlation (XC) functional. In the present work we demonstrate that this approximation lies at the heart of the failure to capture the magnetic structure; local spin density approximation (LSDA)\cite{lsda} and generalized gradient approximation (GGA)\cite{gga} like functionals generate magnetic fields, ({\bf B$_{XC}$}), which have a large source term leading to magnetic mono-poles (i.e. $\boldsymbol{\nabla} \cdot {\bf B}_{XC} \neq 0 $). The presence of this source term in turn leads to large discrepancy in calculated magnetic ground-state and experimental data. Removal of this source-term, by using recently developed source-free XC functional\cite{ss}, reproduces the experimentally observed non-collinear magnetic state (both in terms of the canting angle as well as the magnitude of V moments) in all the three spinel vanadates. Furthermore, we also demonstrate that on-site Coulomb correlation $U$, used to treat strong-correlations in materials via LSDA+$U$ like approach, has the effect of increasing this unphysical source-term, removal of which makes LSDA/GGA+$U$ method highly accurate for the materials under consideration.

The ground state DFT calculations were carried out within full potential linearized augmented plane wave (LAPW) as implemented in the ELK code \cite{elk}. All calculations were performed in the presence of spin-orbit coupling term in the Hamiltonian. A {\bf k}-point grid of $8 \times 8 \times 6$ was used. The exchange-correlation effect were treated using the local spin density approximation (LSDA) and  LSDA+$U$ functionals. A fully unconstrained minimization was performed; a random magnetic field was applied to break the symmetry and subsequently reduced to zero over self-consistent cycle. In this way the self-consistent magnetization is not biased by the initial guess of the magnetization density, which is treated as a unconstrained vector field. The structural parameters of (Mn, Fe, Co)V$_{2}$O$_{4}$ were taken from experiments \cite{nii2, ishi}.

\begin{figure}[!ht]
\centering
\includegraphics[width=9cm]{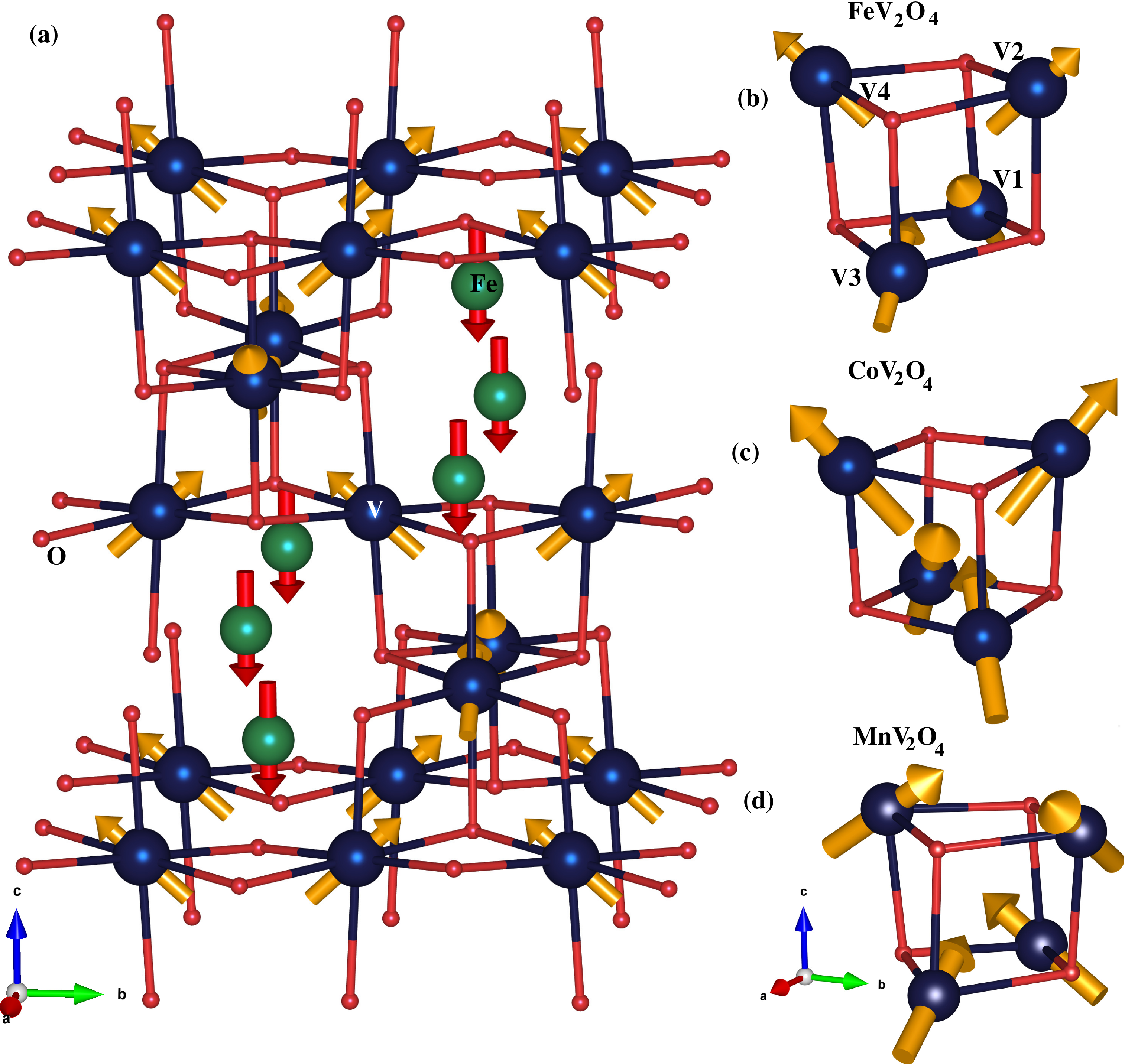}
\caption{The non-collinear magnetic structure obtained using LSDA$_{SF}$+U with U=2.7eV shown (a) within the unit cell for FeV$_2$O$_4$ ; within V$_4$O$_4$ cube for (b) FeV$_2$O$_4$, (c) CoV$_2$O$_4$ and (d) MnV$_2$O$_4$.}
\end{figure}
{\it Spin-texture:} 
In the insulating non-collinear magnetic ground-state observed experimentally all three materials\cite{garlea,chung,mac,ishi,koborinai} possess the A(A=Fe, Co, Mn)-site spin moments aligned along the $c$-axis, while the V moments significantly cant away from the $c$-axis (by an angle of up to\cite{garlea} $65^{\circ}$). In FeV$_2$O$_4$ and CoV$_2$O$_4$, the V-moments form
a structure known as ``two-in-two-out'' structure in each V$_4$O$_4$ cube\cite{mac,jhlee} whereas in MnV$_2$O$_4$ the observed structure is
somewhat more complex\cite{garlea}. 
  
In contradiction to experiments, DFT calculations using LSDA show a metallic collinear ground-state. Since vanadates are strongly correlated insulators  \cite{blanco,slee} adding an on-site Coulomb repulsion by using the LSDA+$U$ method, as expected, opens a gap. However, the magnetic ground-state stays collinear ferrimagnet \cite{sarkar,raman,pardo,dibyendu,ssarkar,nanguneri,tm}, a situation that cannot be improved by changing the functional from LSDA+$U$ to GGA+$U$ or meta-GGA. Adding spin-orbit coupling to the Hamiltonian introduces a weak non-collinearity. 

In order to understand the reason behind this profound discrepancy between theory and experiments we examine the approximate XC functionals used. It has been shown before that for materials such as Fe Pnictides the incorrect magnetic ground-state can be attributed to the unphysical source term in the LSDA (and GGA) XC magnetic fields, removal of which, via source-free XC functional, results in agreement with experiments\cite{ss}. In the present case the problem is more complex in that not only the magnitude (as in the case of Fe-Pnictides), but also the direction of the local moments obtained using LSDA+$U$ functional are incorrect. Whether the source term in LSDA+$U$ magnetic field is also responsible for the incorrect spin texture in these materials remains to be seen. 

We employ the source-free LSDA+$U$ functional (hereafter denoted by LSDA$_{SF}$+$U$) to perform a fully unconstrained optimization of magnetization density (both the direction and magnitude). In agreement with experiments we find a non-collinear magnetic ground state for all three compounds. Most importantly, a single value of $U$ (i.e. $U$=2.7 eV acting on the V-atoms) is required to reproduce the experimentally observed diverse non-collinear magnetic ground states of all the three compounds: both the ``two-in-two-out''\cite{mac,jhlee} spin arrangements of FeV$_2$O$_4$ and CoV$_2$O$_4$ (Fig.~1a and 1b) and the complex spin-texture in  MnV$_2$O$_4$\cite{mac,jhlee} (Fig. 1c) are perfectly captured.

\begin{figure}[!ht]
\centering
 \includegraphics[width=9.0cm]{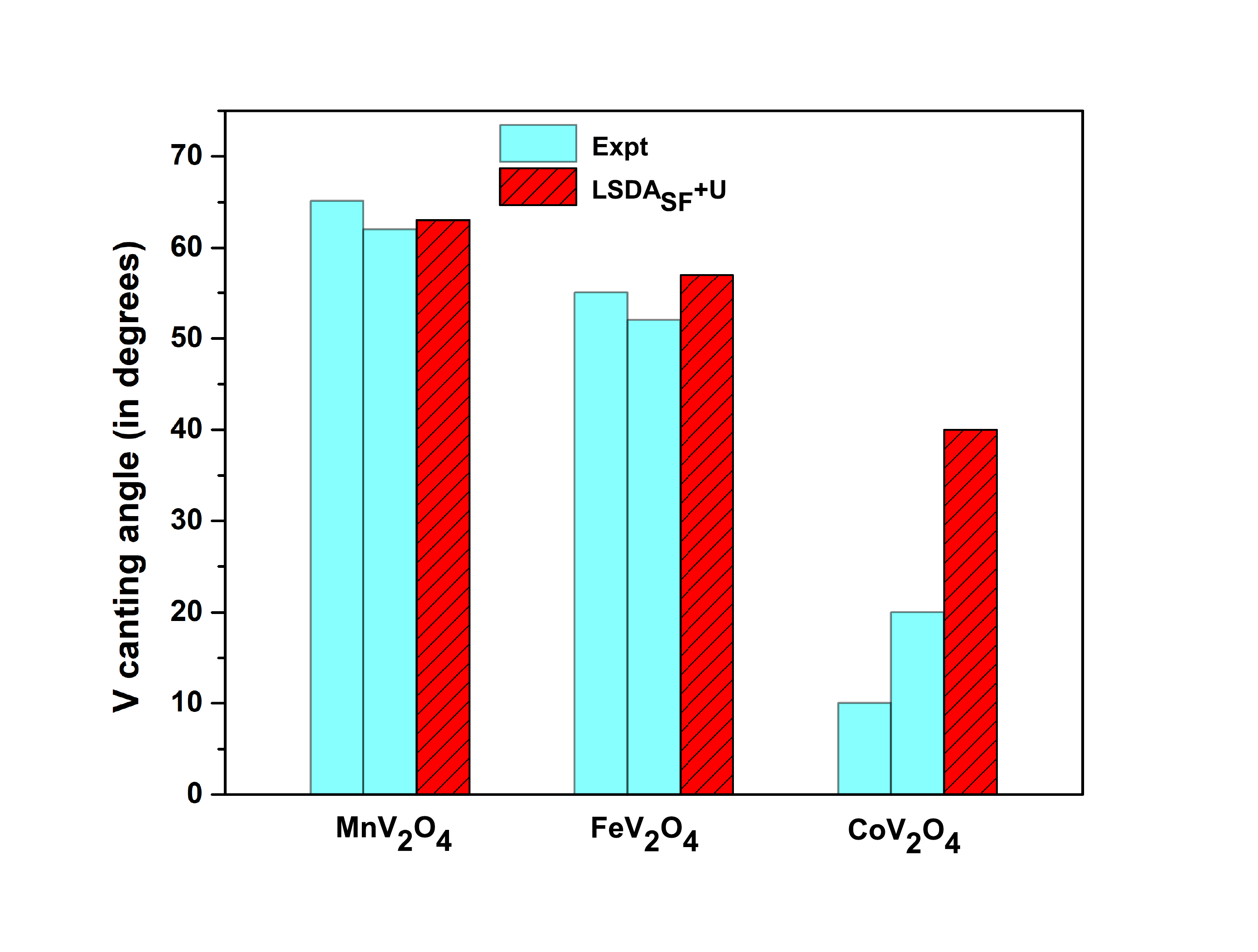}
 \caption{Canting angle of V moments with respect to the $c$-axis calculated with source-free LSDA+$U$ (red) and compared with the corresponding experimental values\cite{garlea,lzhang,mac,ishi,koborinai,nakamura} (cyan) for MnV$_{2}$O$_{4}$, FeV$_{2}$O$_{4}$ and CoV$_{2}$O$_{4}$. The on-site Coulomb repulsion $U$ is equal to 2.7 eV for all materials. Note that employment of the standard LSDA functional results in a grossly wrong value of the canting angle of 0$^\circ$ which increases to 9$^\circ$ upon inclusion of the spin-orbit coupling term.}
\end{figure}

As for the value of the angle between the V and A site spins, consistent with the experimental observations our results show that the A site moment is collinear with the $c$-axis in presence or absence of the source term in the functional. On the other hand, the canting angle of the V spins is highly functional dependent in that LSDA+$U$ functional, in presence of spin-orbit coupling, leads to a small canting angle of $19^{\circ}$. These results are unlike experiments which show that the canting angle is much smaller in CoV$_2$O$_4$ than in MnV$_2$O$_4$ or FeV$_2$O$_4$. Removal of the source term from this functional has a dramatic effect on the canting angle (see Fig. 2); for MnV$_2$O$_4$ and FeV$_2$O$_4$ the agreement with experiments is excellent but for CoV$_2$O$_4$ the results overshoot slightly. However, consistent with the experimental trend \cite{koborinai,mac,garlea,lzhang,nakamura} we find that the canting angle is smaller in CoV$_2$O$_4$ than in MnV$_2$O$_4$ or FeV$_2$O$_4$.



\begin{table}[]
\centering
\caption{Magnetic moments (in $\mu_{B}$) per A-site atom.}
\
\renewcommand{\arraystretch}{1.5}
\label{tab1}
\scalebox{0.7}{
\begin{tabular}{||c|c|c|c|c|c||}
\hline
\textbf{Vanadate} & \textbf{Expt.} & \textbf{LSDA} & \textbf{$LSDA_{SF}$} & \textbf{LSDA+U} & \textbf{$LSDA_{SF}+U$} \\
		 & & & & \textbf{($U$=2.7 eV)} &\textbf{($U$=2.7 eV)} \\
\hline\hline
\textbf{$MnV_{2}O_{4}$} & 4.2\cite{garlea}, 4.11\cite{oka} & 4.02 & 3.98 & 4.11 & 4.10 \\
\hline
\textbf{$FeV_{2}O_{4}$} & 4\cite{mac} & 3.22 & 3.15 & 3.42 & 3.44 \\
\hline
\textbf{$CoV_{2}O_{4}$} & 2.46\cite{nonaka} & 2 & 2.08 & 2.45 & 2.41 \\
\hline\hline
\end{tabular}}
\end{table}

\begin{figure}[!ht]
\centering
\includegraphics[width=9.0cm]{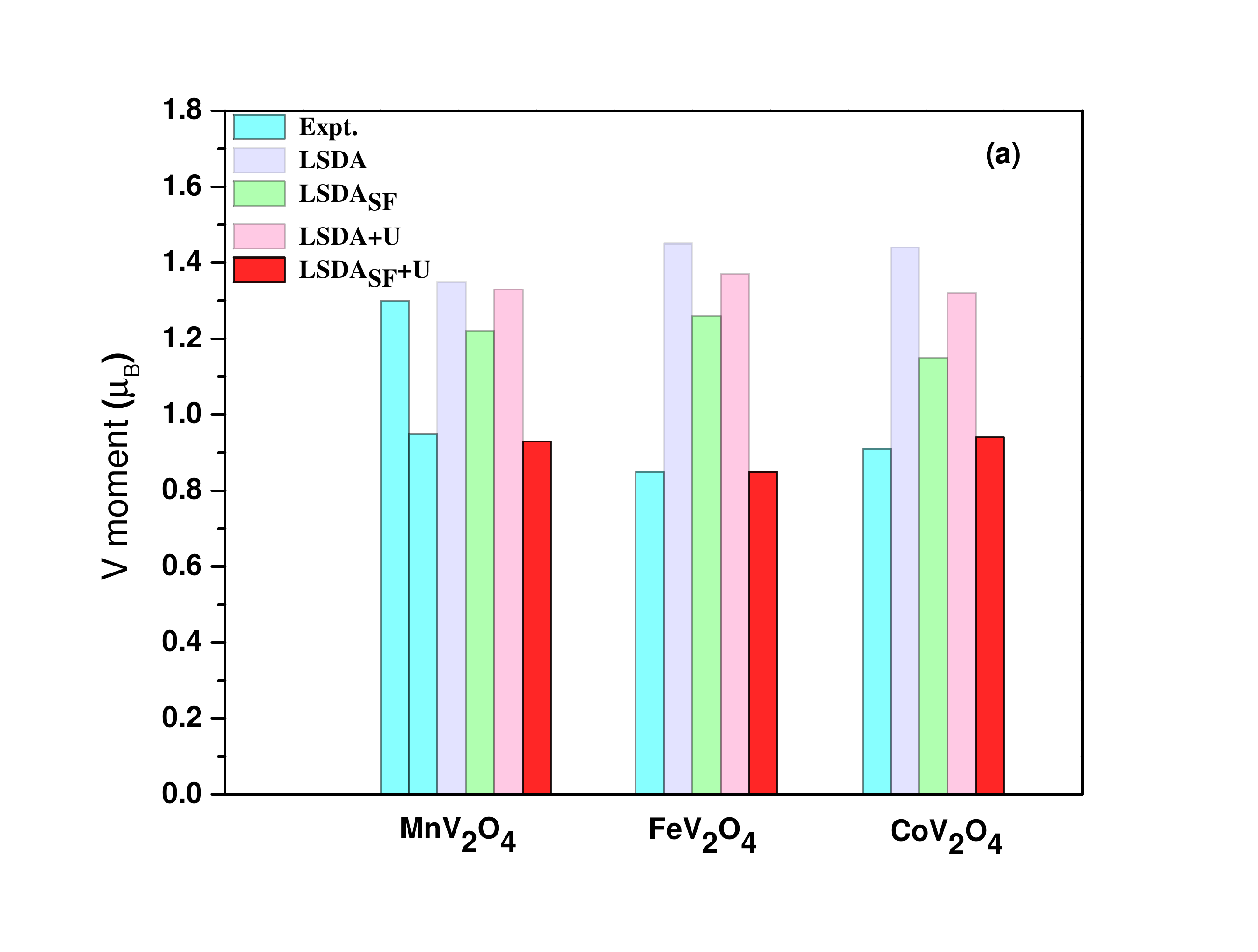} \\
\includegraphics[width=7cm]{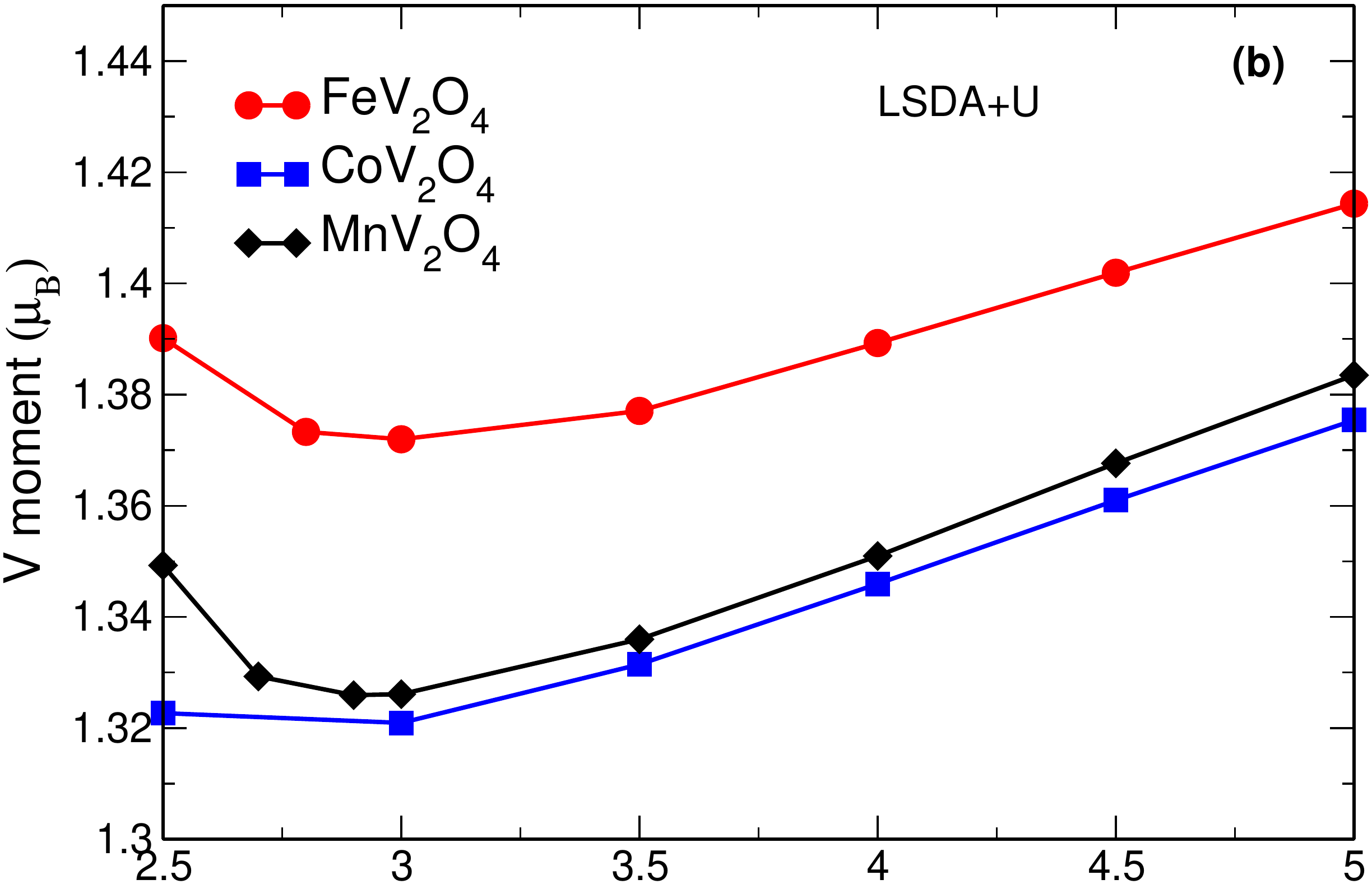} \\
\includegraphics[width=7cm]{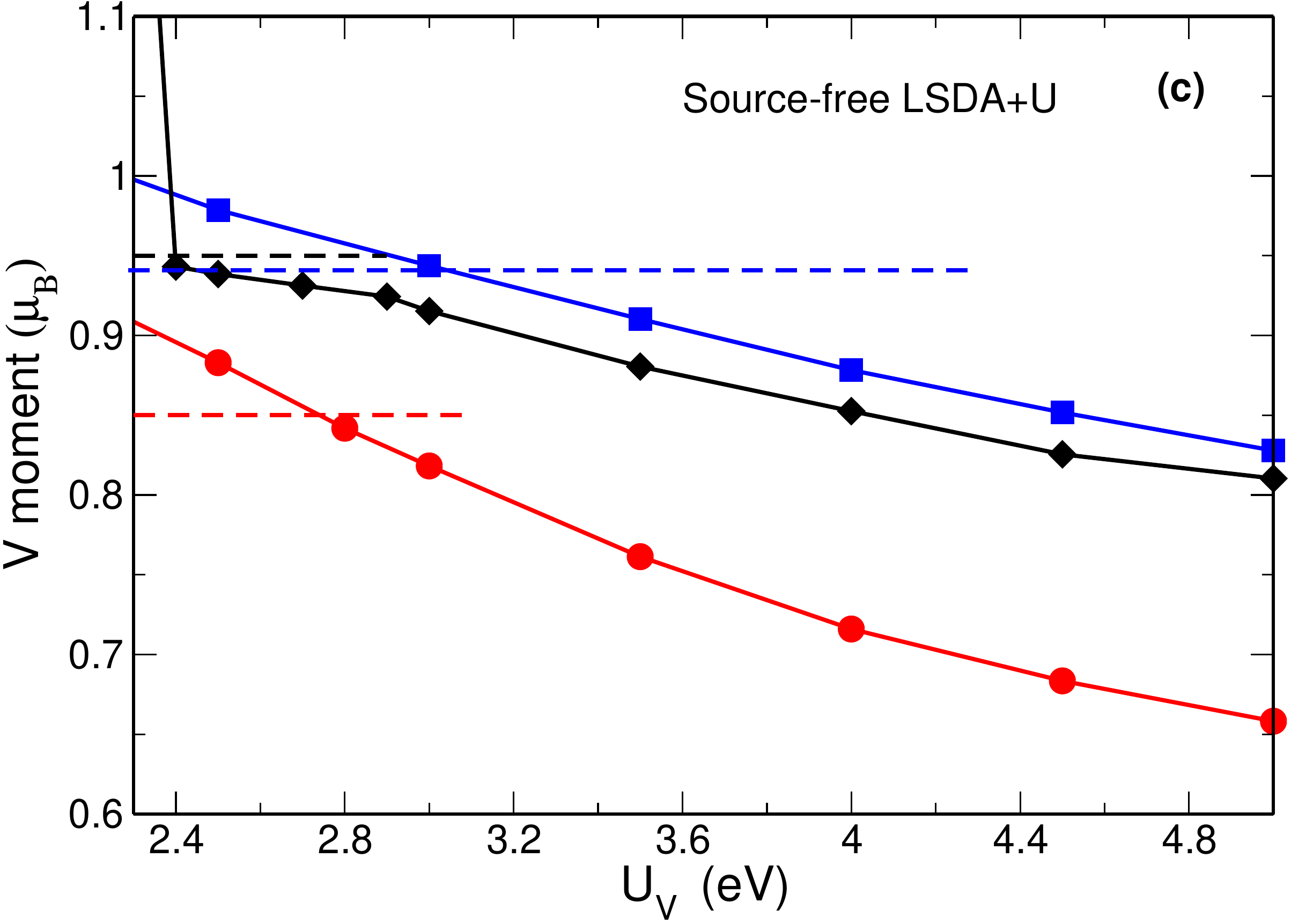} 
\caption{ (a) V magnetic moment calculated using LSDA (light blue), LSDA$_{SF}$ (green), LSDA+$U$ (pink) and LSDA+$U_{SF}$ (red) compared with corresponding experimental (cyan) values \cite{garlea, jhlee, mac, ishi, nonaka} for $MnV_{2}O_{4}$, $FeV_{2}O_{4}$ and $CoV_{2}O_{4}$. The on-site Hubbard $U$ parameter is set equal to 2.7~eV for all materials. 
Variation of total spin magnetic moment of  V  as a function of this on-site Hubbard $U$ using (b) LSDA+$U$ and (c) source-free LSDA+$U$ V$(m_{V})$ functionals.}
\end{figure}

{\it Magnitude of the moment:} 
As discussed in the introduction, experimental measurements via neutron diffraction \cite{garlea,jhlee,mac}, XMCD \cite{nonaka} etc. on spinel vanadates report a small moment on the V atoms in all these compounds;  1.3 $\mu_B$ in MnV$_2$O$_4$, 0.85 $\mu_B$ in FeV$_2$O$_4$, 0.9 $\mu_B$ in CoV$_2$O$_4$. This moment is much smaller than 2$\mu_B$, the expected value for a V$^{3+}$ state. Furthermore, XMCD measurements performed on MnV$_2$O$_4$ and FeV$_2$O$_4$ reveal a very small value of the orbital moment \cite{oka} indicating the V moment is primarily spin in character \cite{matsura}, and so the cancellation of the spin-moment by the orbital moment cannot be the reason behind this reduction in the moment. As for the moment on the A-site is concerned, all experiments report a large moment \cite{garlea,oka,mac,nonaka}. 

In contraction to these experiments, and in agreement with previous \emph{ab-initio} work\cite{sarkar, dibyendu}, we find that DFT calculations performed using the LSDA and LSDA+$U$ functionals show a large moment on V atoms with percentage deviation of up to 40\% from experiment (see Fig. 3a).
Interestingly,  there does not exist a value of $U$ for which the correct moment on the V atoms can be obtained (see Fig. 3b). Use of LSDA$_{SF}$ and LSDA$_{SF}$+$U$ with $U$=2.7 eV, the value of $U$ that gives correct spin texture, remarkably, also leads to the value of the V moment in close agreement with experiments with the worst error only a 2\% deviation (see Fig. 3a). The moment on the A site is well described by LSDA/LSDA+$U$ and their source-free counterparts (see Table I). 
Thus the source free LSDA+$U$ functional provides complete description of the ground-state of all three spinel vanadates with a single value of $U$. 

A study of the magnitude of the V moment as a function of $U$ leads to a striking observation; increase of $U$ within LSDA+$U$ functional generates, as expected, increased on-site localization of charge and an increased local moment on the V atoms (see Fig. 3(b)). A consequence of this is that there does not exist a value of $U$ for which the correct value of V moment is obtained.  
However, the V moment calculated by excluding the source term from the LSDA+$U$ functional shows exactly the opposite trend: the V moment \emph{decreases} with $U$ (see Fig. 3(c)). This is a counter intuitive yet explainable trend-- as the value of $U$ increases the source term in the XC magnetic field also increases. The removal of this source term then has a significant effect on the magnetization density leading to a decrease in the V moment as a function of increasing $U$. This suggests that in treating the magnetic ground state of strongly correlated materials within the LSDA+$U$ framework, varying $U$ can in an uncontrolled way also alter the unphysical source term in ${\bf B}_{XC}$. Thus it would appear imperative in treating strong correlated magnetic ground states that ${\bf B}_{XC}$ should be source free.


In conclusion, we have investigated the magnetic ground-state for spinel vanadates AV$_2$O$_4$ (A=Mn, Fe and Co) using density functional theory. In doing so we find that the well known failure of all traditional XC functionals (LSDA, GGA, LSDA+$U$, meta GGA) to reproduce the experimentally observed magnitude of local moments and non-collinear spin texture arises from the presence of a large source term in the magnetic field generated by these functionals. Most strikingly, we find that this source-term increases on increasing the value of $U$.
Removing this unphysical source-term from LSDA+$U$ functional we find results in a perfect description of the ground-state magnetism of these materials. Most importantly we find that for all three materials we needed the same value of $U$ for this correct description. This is a great improvement over traditional LSDA+$U$ approach where there does not exist a single value of $U$ which gives the correct ground-state for these materials.

This work is supported by DST-DAAD project (grant no INT/FRG/DAAD/P-16/2018) funded by DST(India) and DAAD (Germany). JK would like to acknowledge MHRD(INDIA) research fellowship. Sharma expresses gratitude to the DFG for funding through TRR227 (project A04).

\end{document}